\documentclass[submission]{eptcs}
\providecommand{\event}{REFINE 2011} % Name of the event you are submitting to
\usepackage{breakurl}             % Not needed if you use pdflatex only.

\usepackage{amsmath}
\usepackage{circus}
\usepackage{graphicx}
\usepackage{stateflow}

\renewcommand{\circblockbegin}{\left(\begin{array}{l}}
\renewcommand{\circblockend}{\end{array}\right)}

\title{Refinement-based verification of sequential\\implementations of Stateflow charts}
\author{Alvaro Miyazawa
\institute{Department of Computer Science\\University of York}
\email{alvarohm@cs.york.ac.uk}
\and
Ana Cavalcanti
\institute{Department of Computer Science\\University of York}
\email{ana.cavalcanti@cs.york.ac.uk}
}
\def\titlerunning{Refinement-based verification of Stateflow charts}
\def\authorrunning{A. Miyazawa, A. Cavalcanti}

\begin{document}

\maketitle

\begin{abstract}
Simulink/Stateflow charts are widely used in industry for the
specification of control systems, which are often safety-critical. 
This suggests a need for a formal treatment of such models. In previous work,
we have proposed a technique for automatic generation of formal models
of Stateflow blocks to support refinement-based reasoning. In this article, we present a refinement strategy that supports
the verification of automatically generated sequential C implementations
of Stateflow charts.
In particular, we discuss how this strategy can be specialised to
take advantage of architectural features in order to allow a higher level
of automation.
\end{abstract}

\section{Introduction}

MATLAB Simulink~\cite{Simulink} is a graphical notation widely used in the automotive and avionics industries; it
supports the specification of control systems in a level of abstraction convenient for engineers.
A~Simulink diagram consists of blocks and wires connecting the inputs
and outputs of the blocks.% The execution of the diagram consists of
%a series of steps, in which all the diagram's blocks are executed in a
%particular order dependent on the wiring.

Stateflow~\cite{Stateflow} is an extension of Simulink that supports the specification of state transition systems, providing a new Simulink block, namely, a Stateflow chart.
It is a variant of Statecharts~\cite{Harel1987}, which extends standard state-transition systems by introducing new features,
such as hierarchy and parallelism.

While Simulink diagrams are typically used to specify aspects of a system that can be modelled by differential
equations relating inputs and outputs, Stateflow charts usually model the control aspects. There is a wide range of
tools that support Simulink and Stateflow. These include a simulation and analysis tool, a verification and validation tool, a code generator
and a prototyping tool~\cite{Simulink,Stateflow,RealTimeWorkshop}.

The extensive use of Simulink/Stateflow in the development of safety-critical systems, associated with
certification standards~\cite{BS2002, DO178b} that recommend the use of formal methods for the specification, design,
development and verification of software, makes a formal treatment of these notations extremely useful.

\begin{figure}
\centering
\includegraphics[width=.9\textwidth]{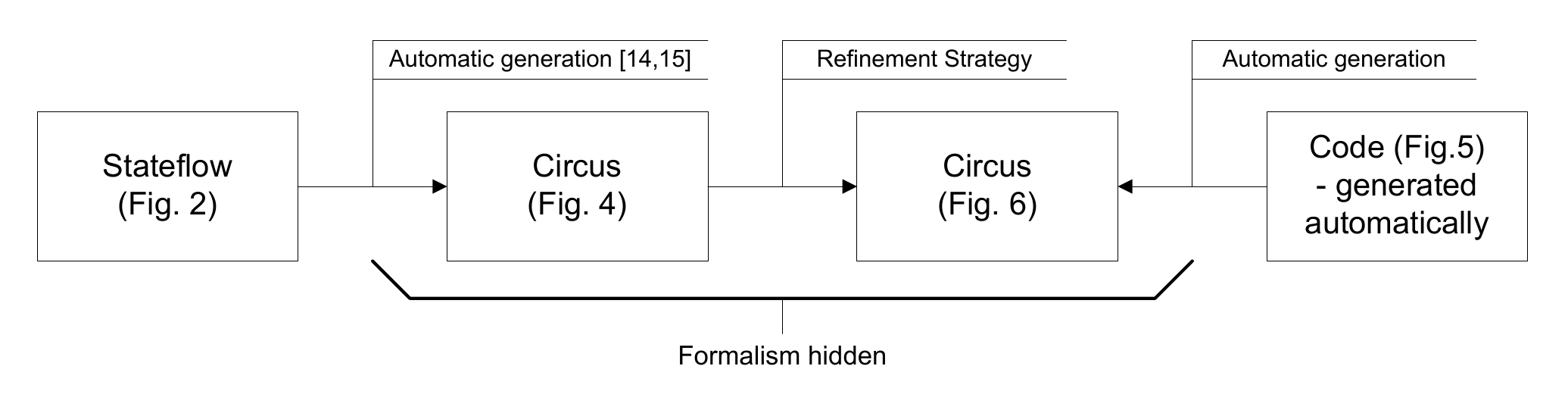}
\caption{An approach for the verification of implementations of Stateflow charts.}
\label{fig:approach}
\end{figure}

We are concerned with the assessment of the correctness of implementations of Stateflow charts.
A frequent approach to this problem is based on the verification of automatic code generators~\cite{Caspi2003,Toom2008,Lublinerman2009}. 
%This has been frequently dealt with by approaches based
%on the verification of automatic code generators~\cite{Caspi2003,Toom2008,Lublinerman2009}. 
We propose an orthogonal approach based on the verification of implementations with respect to a model
of the chart. An overview of our approach is given in Figure~\ref{fig:approach}.
%We propose an approach that is orthogonal to the verification of code generators; it
%is based on the verification of implementations with respect to a model of the chart. Figure~\ref{fig:approach} depicts our approach.% An overview of our approach is given in Figure~\ref{fig:approach}.

This approach consists of deriving formal models of a Stateflow chart and its implementation, and
applying the refinement calculus to check the correctness of the model of the implementation with respect to the model of the chart.
This is particularly suited for situations where automatically generated code is not applicable
or convenient, for instance, in situations where hardware and performance requirements require changes in the generated
code. Moreover, Simulink and Stateflow are frequently updated, and these updates can have
a heavy impact on the cost of the verification of any code generator.

In \cite{Miyazawa2011a}, we propose an operational model of Stateflow charts, provide translation rules for deriving such models, and discuss
a tool that automatically generates the model of a Stateflow chart. The model of a Stateflow chart is formed by the composition of two processes:
the first models the general semantics of Stateflow~\cite{Stateflow}, and the second models specific aspects of a chart.
The model of the semantics of Stateflow chart is structured in a way that facilitates the inspection and comparison to the informal semantics found in~\cite{Stateflow},
as no formal analysis can be made because the semantics of Stateflow is available in an informal way~\cite{Stateflow} or is hidden in the simulation tool.

The refinement calculus is enough for the purpose of verifying such models. However, the expertise required for such verification is often not available.
Moreover, the complexity of Stateflow and the size of real charts potentially renders the manual application of the refinement calculus infeasible. We aim
in our approach to hide as much of the formalism as possible, to allow it to be used in real scenarios by engineers and programmers. For that to be achieved, we must
provide means for the refinement to be established at least in a semi-automatic way.% This paper is a first attempt at solving this problem.

We propose a verification strategy for sequential automatically generated implementations of discrete-time Stateflow charts with respect to
models of Stateflow constructed (automatically) as described in \cite{Miyazawa2011a}. This technique is closely related to that
proposed in~\cite{Cavalcanti2011} for verification of implementations of Simulink diagrams. Our work extends those results
to cover a larger class of diagrams and implementations.

The implementations that we consider are those that follow the architectural pattern employed by the code generator provided
by MATLAB. There are other code generators \cite{Scaife2004,Toom2008}, but as far as we know, they all cover
a limited subset of the Stateflow notation. Fixing the architecture of the implementation allows us to specialise the details of the
strategy to increase its level of automation.

Our models for Stateflow charts are specified in \Circus~\cite{Woodcock2002}, a formal notation that integrates Z~\cite{Woodcock1996},
CSP~\cite{Roscoe1998}, Dijkstra's language of guarded commands~\cite{Dijkstra1975}, and the refinement calculus~\cite{Morgan1994}.
These models are particularly adequate for refinement-based verification techniques. Our technique uses the \Circus\ refinement laws to provide
a tactic of refinement that can be used to prove the correctness of an implementation in a highly automated way. Soundness of the technique
stems from soundness of the laws.

This article is structured as follows. Section~\ref{sec:background} introduces the background material necessary for the presentation of our strategy.
Section~\ref{sec:impl} discusses the architecture of automatically generated implementations of Stateflow charts and provides general guidelines for
deriving \Circus~models of these implementations. Section~\ref{sec:strategy} describes our refinement strategy for the verification of implementations
of Stateflow charts. Section~\ref{sec:conclusion} assesses our contributions, examines related work, and discusses directions for future developments.

\section{Background material}
\label{sec:background}

In this section, we introduce the Stateflow and \Circus~notations, and our formal models of Stateflow~charts.%modelling approach for Stateflow.

\subsection{Stateflow charts}

\begin{figure}
\centering
\includegraphics[trim=0cm 9cm 0cm 8.5cm,width=.7\textwidth,clip=true]{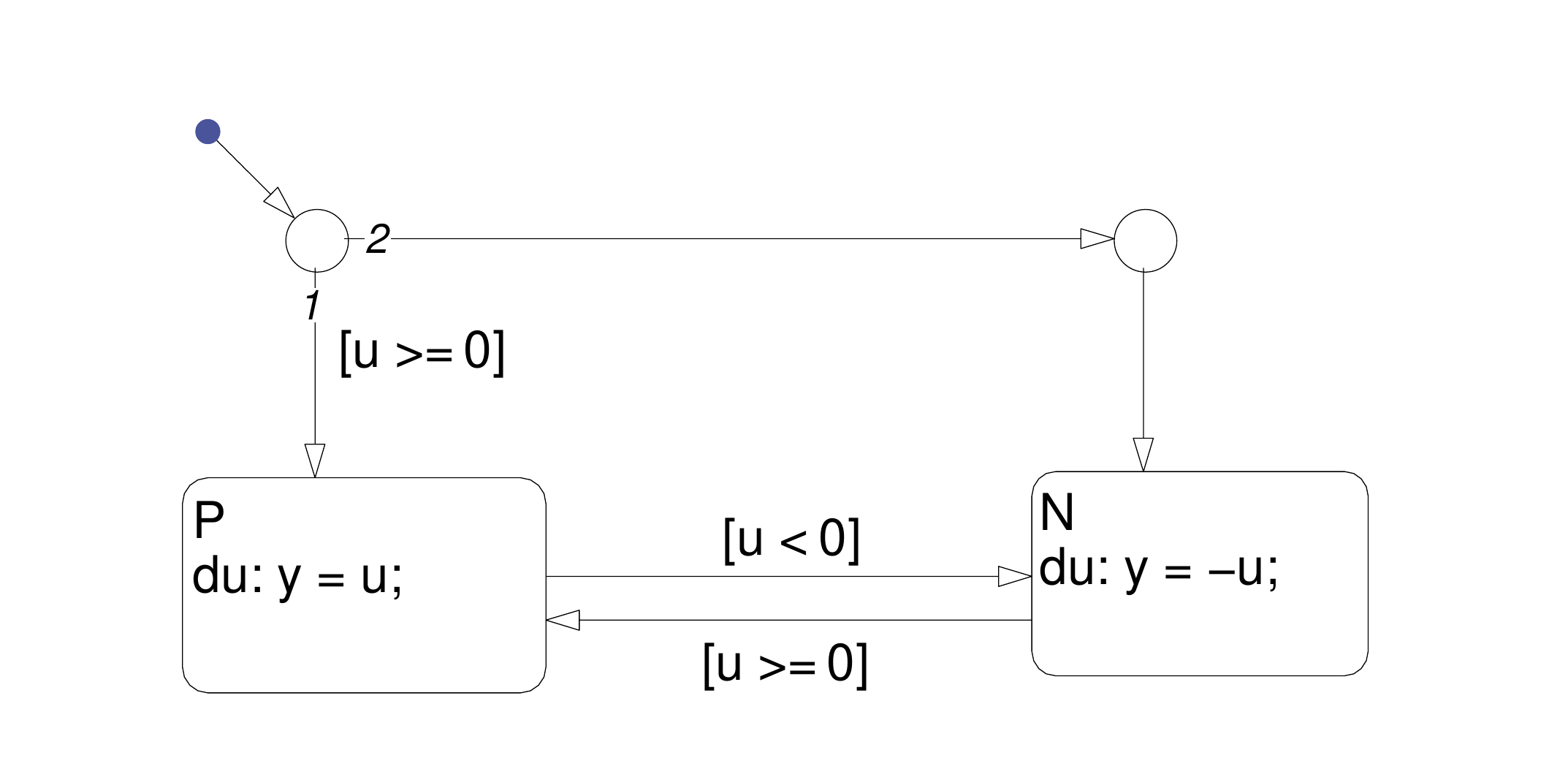}
\caption{Absolute value chart.}
\label{fig:example}
\end{figure}

Figure~\ref{fig:example} shows our running example: a Stateflow chart adapted from an example supplied with the tool.
The chart has one input variable (\texttt{u}) and one output variable (\texttt{y}); it outputs in \texttt{y} the absolute value of \texttt{u}.

A Stateflow chart is built from a series of components, such as states, transitions, junctions, data and events.
States are represented by rectangles with round corners; in our example, the boxes marked with \texttt{P} and \texttt{N} are states. States, as well as charts, can have substates, which are arranged in a sequential or parallel decomposition. A state with a sequential decomposition has at most one substate active at any given time, while a state with a parallel decomposition has all of its substates active or inactive at once.

A state has a set of actions associated with it, namely, entry, during, exit, on, and binding actions. Entry, during and exit actions are executed when the state is entered, executed, and exited, respectively; on actions are executed in the same situations as during actions, with the additional requirement that a particular event is being processed; and binding actions bind a particular event or data to the state. In our example, both \texttt{P} and \texttt{N} have a during action. In \texttt{P}, \texttt{u} is assigned to \texttt{y}, and in \texttt{N}, \texttt{-u} is assigned to \texttt{y}.

Two states (within a state or chart with sequential decomposition) can be connected by one or more transitions; they are indicated by arrows and can be guarded by events and conditions. There are two types of actions associated with a transition: condition and transition actions. Condition actions are executed when the guard of the transition (event and condition) is true, and transition actions are executed when the transition leads to a state being exited.

Transitions are classified according to the relative position between its source and target states; inner transitions have the target state as a substate of the source state, and outer transitions do not. There is a special type of transition, called default transition, that has no source; it is used to indicate the default path to be taken when a state or chart is first entered. In our example, we have one default transition, and five outer transitions. Three of the transitions are guarded by a condition: \texttt{u>=0}, \texttt{u<0}, or \texttt{u>=0}.

A transition path is formed by a series of transitions linked by junctions which are represented by circles. There are two junctions in our example; they form two transition paths. A transition path is completed only when a state is reached by following all the transitions in the path. When a transition path is completed, the source of the path is exited, the transition actions of the path are executed, and the target state is entered. Additionally, there is a special type of junction, called history junction, which records the most recently activated substate of the state that contains it.

In the example in Figure~\ref{fig:example}, initially, the chart is inactive; the first time it is executed, it is activated and one of its two states is entered depending on whether %its input
\texttt{u} is greater than or equal to zero (state \texttt{P}) or not (state \texttt{N}). In the next execution, if the sign of \texttt{u} has changed, a transition takes place from one state to the other. If there is no change, the during action of the active state assigns the absolute value of \texttt{u} to \texttt{y}.

Before presenting the \Circus~model of this chart, we give, in the next section, an overview of \Circus.

\subsection{\Circus}

We present the main \Circus\ features using the example in~Figure~\ref{fig:merge}. It models a parallel sorter that reads a sequence of natural numbers through the channel $in$, and writes on the channel $out$ an ordered version of the input sequence. A detailed presentation of \Circus\ can be found in~\cite{Woodcock2002}.

\begin{figure}[t]
\centering
\begin{minipage}{.8\textwidth}
\begin{circus}
\circchannel in, in1,in2,out: \seq \nat\\
\circprocess Merger \circdef \circbegin\\
\t1\circstate S == [y:\seq\nat]\\
\t1 InitS == [S~' | y' = \langle\rangle]\\
\t1~Merge ~\circdef~ x1, x2:\seq \nat~\circspot~\\ 
\t2\circblockbegin 
\circif \# x1 = 0 \circthen y:=y\cat x2\\
\circelse \# x2 = 0 \circthen y:=y \cat x1\\
\circelse \# x1 \neq 0 \land \# x2 \neq 0 \circthen\\
\circblockbegin
\circif head~x1 \leq head~x2 \circthen y:=y\cat \langle head~x1 \rangle \circseq Merge(tail~x1,x2)\\
\circelse head~x1 > head~x2 \circthen y:=y\cat \langle head~x2 \rangle \circseq Merge(x1,tail~x2)\\
\circfi
\circblockend\\
\circfi
\circblockend\\
\t1\circspot 
InitS \circseq in1?x1 \then in2?x2 \then Merge(x1,x2) \circseq out!y \then \Skip\\
\circend\\
\circprocess SplitSorter \circdef \ldots\\
\circprocess ParallelSorter \circdef \left(\begin{array}{c}
SplitSorter\\
\lpar \lchanset in1,in2 \rchanset \rpar\\
Merger\end{array}\right)\circhide \lchanset in1,in2 \rchanset
\end{circus}
\end{minipage}
\caption{The $ParallelSort$ specification.}
\label{fig:merge}
\end{figure}

A \Circus~specification is a sequence of paragraphs:~Z paragraphs~(axiomatic definitions, schemas, and so on), channel and channel set declarations, and process definitions. The first paragraph of our example defines four channels $in$, $in1$, $in2$, and $out$, which communicate sequences of natural numbers, that is, elements of the type $\seq \nat$. The second paragraph is a basic process definition. It provides the name of the process ($Merger$), the definition of its state using a schema $S$, an action $InitS$ defined by an operation schema, an action $Merge$, and a main action (after a $\circspot$), which defines, using the previously defined actions, the overall behaviour of the process.

In general, \Circus\ actions are written using a mixture of Z and CSP constructs, and guarded commands. In our example, the main action initialises the state using $InitS$, reads a value $x1$ through the channel $in1$, reads a value $x2$ through $in2$, calls $Merge$ with the values $x1$ and $x2$ as parameters, and outputs the state variable $y$ through $out$.

The schema $S$ has only one component $y$ of type $\seq \nat$, that is, the set of sequences of natural numbers. The schema $InitS$ specifies an initialisation operation over $S$ that sets $y$ to the empty sequence ($\langle \rangle$). Like in Z, we use $y'$ to refer to the value of $y$ after the operation.

The action $Merge$ takes two sequences $x1$ and $x2$ of natural numbers, and appends them to the state variable $y$, so that if both input sequences are ordered, the final sequence in $y$ is also ordered. The specification of $Merge$ uses a conditional and assignments from the guarded commands language. If one of the sequences is empty, the non-empty sequence is appended. When both sequences are not empty, $Merge$ compares the first element of each sequence ($head~x1 \leq head~x2$), appending the smallest of them to $y$, and recursively calls $Merge$ on the rest of the sequence that had the smallest element ($tail~x1$ or $tail~x2$), and the whole of the other sequence.

Processes encapsulate their state and interact with other processes through channels. The usual CSP operators can be used to combine processes. The fourth paragraph in Figure~\ref{fig:merge} defines the process $ParallelSorter$ as the parallel composition of the processes $SplitSorter$ and $Merge$, communicating over the channels $in1, in2$. The process $SplitSorter$ in the third paragraph is omitted; it splits a sequence of natural numbers in two, sorts each sequence in parallel, and outputs them through channels $in1$ and $in2$. In the definition of $ParallelSorter$, the channels $in1$ and $in2$ are hidden, thus yielding a process whose interface contains only the channels $in$ and $out$.

In the next section, we have another example of a process; the model of the chart in Figure~\ref{fig:example}.

\subsection{A formal model of Stateflow charts}
\label{sec:formal-model}

In this section, we describe the \Circus~operational models of Stateflow
charts that we can generate automatically. In these models, 
the execution of one step of the chart is initiated by
reading inputs, and concluded by writing outputs, and synchronising
on a channel called $end\_cycle$. A more detailed description
can be found in \cite{Miyazawa2011,Miyazawa2011a}.

\begin{figure}[t]
\centering
\includegraphics[width=.8\textwidth]{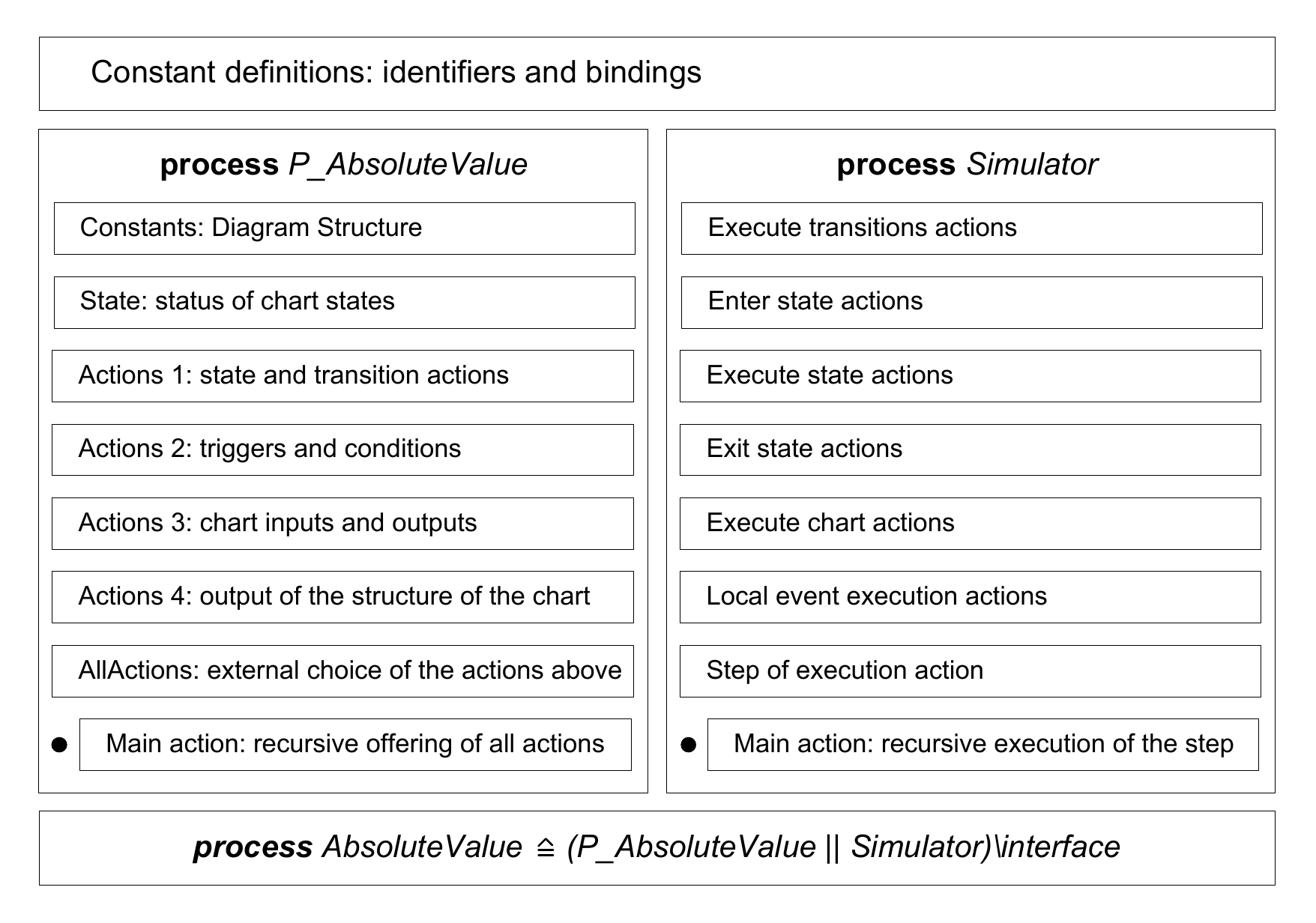}
\caption{Structure of the model of the chart in Figure~\ref{fig:example}}
\label{fig:model}
\end{figure}

Our models consist of two \Circus~processes in parallel. The
first, $Simulator$, represents the simulator, and is the same for every chart.
The second, the chart process, represents a particular chart.
The simulator and the chart processes communicate over the channels in the set $interface$ plus
the channel $end\_cycle$, with the channels in $interface$ hidden.
Figure~\ref{fig:model} shows the structure of the automatically
generated model of the chart in Figure~\ref{fig:example}.

%Figure~\ref{fig:model} shows part of the chart process $P\_AbsoluteValue$ for the example in
%Figure~\ref{fig:example}, and how it is composed with the simulator process
%to define the process $AbsoluteValue$ that models the diagram.
%This process can communicate with the environment only through the channels~$i\_u$,
%$o\_y$, and $end\_cycle$.

The chart process $P\_AbsoluteValue$ uses a data model that defines the state, transition, and junction identifiers, as well as the states, transitions, and junctions as bindings of specific schemas. These are constants that capture information about the structure of the chart. They are represented by the first rectangle in Figure~\ref{fig:model}.
% In our example, they include, for instance, $s\_P$ and $s\_N$, the identifiers of the states \texttt{P} and $\texttt{N}$, and $S\_P$, a binding that records information about \texttt{P} (its identifier, the fact that is has no history junctions, and so on).
These constants are collected in four other constants defined within the chart process:~$identifier$, $states$, $transitions$, and $junctions$. The constant $identifier$ records the identifier of the chart and $states$, $transitions$, and $junctions$ are partial functions that map identifiers to the corresponding binding. These constants are declared using a schema $StateflowChart$ whose definition is omitted in Figure~\ref{fig:model}. Their values are fixed in the process chart.% For example, in Figure~\ref{fig:model}, we define $identifier$ to be $c\_AbsoluteValue$.

Next, the chart process defines a series of schemas that specify components of the state and corresponding initialisation operations. Information about which states are active and which states are recorded in the history junctions is recorded in the schema $SimulationData$, and chart variables are recorded in the schema $SimulationInstance$. These schemas are conjoined to define the schema $State$ that specifies the state of the process. We adopt the convention of prefixing a $v\_$ to the name of the chart variable to clarify the nature of the name.

Next, the chart process defines a series of \Circus~actions that can be divided into four groups: actions that correspond to state and transition actions, actions that correspond to calculation of triggers and conditions, actions that read inputs and write outputs, and actions that output the structure of the chart.
%
%The first group contains actions such as $entryaction\_P$ and $duringaction\_P$ in Figure~\ref{fig:model}; they are prefixed by a communication over an appropriate channel (for example, $executeentryaction$ for entry actions) of the identifier of the state (or transition) followed by a \Circus~action that encodes the Stateflow action. Since there is no entry action associated with the state $P$, its action is simply the communication followed by the action $\Skip$.
%
%The second group contains actions such as $condition\_P\_N$ and $trigger\_P\_N$ in Figure~\ref{fig:model}; they encode the verification of the condition and of the trigger of the transition between the states $P$ and $N$. The action $condition\_P\_N$ is defined as a conditional whose guards are the condition itself and its negation, and whose corresponding actions are a communication on the channel $evaluatecondition$ that communicates the transition identifier and a boolean value corresponding to the evaluation of the guard. Since our example has no events, the action $trigger\_P\_N$ always communicates $\true$ through the channel $result$, indicating that this transition has no trigger.
%
%The third group contains two actions $Inputs$ and $Outputs$, which are a prefixing on the events $read\_inputs$ and $write\_outputs$, respectively. Finally, the fourth group contains actions such as $getstate$, which takes a state identifier and communicates the binding corresponding to it. 
%
All these actions are grouped to define $AllActions$, which is used in the main action as shown in Figure~\ref{fig:chart-main-action}.

\begin{figure}
\centering
\hspace{-.8cm}
\begin{minipage}{\textwidth}
\begin{circus}
\circspot \lschexpract InitState \rschexpract \circseq
\circmu X \circspot \circblockbegin\circblockbegin
\circmu Y \circspot
\circblockbegin
((AllActions\circinterrupt(interrupt\_chart\then\Skip))\circseq Y)\\
\extchoice\\
end\_cycle\then\Skip
\circblockend
\circblockend\circseq X\circblockend\\
\circend
\end{circus}
\end{minipage}
\caption{Main action of the chart process shown in Figure~\ref{fig:model}}
\label{fig:chart-main-action}
\end{figure}

The main action of the chart process is shown in Figure~\ref{fig:chart-main-action}; it initialises the state, and recursively offers the actions in $AllActions$, with the additional possibility that any of these actions can be interrupted at any time by a communication over the channel $interrupt\_chart$. The possibility of interruption accounts for the occurrence of early return logic in the chart, that is, the interruption of the execution of the chart brought about by a state inconsistency produced by a local event broadcast. The main action has two nested recursions: the internal one corresponds to the actions that are offered to one particular step of simulation, and can be terminated by a synchronisation over the channel $end\_cycle$. The external recursion corresponds to the recursive execution of simulation steps.

The process $Simulator$ does not have a state; it declares a series of actions that model the execution of transitions, as well as the processes of entering, executing and exiting states. The execution of a chart is then defined in terms of the previous actions. A chart can be executed multiple times in the same time step due to the occurrence of multiple input events or the broadcast of a local event. In the first case, an action encodes the execution of the chart for each active event in the appropriate order and is used to define the step of execution. In the second case, we define an action that captures the occurrence of a broadcast and executes the chart under the appropriate setting. This action is called whenever a state or transition action is executed.
%
%These actions are used in the definition of the step of execution. The case of a local event broadcast is modelled by two actions that capture the occurrence of a broadcast and execute the chart under the appropriate setting. The treatment of a local event is triggered whenever a state or transition action is executed because they potentially generate local event broadcasts.
%
All these actions are combined to define the step of execution of the simulator, which is called recursively in the main action of the process $Simulator$.

In the next section we discuss an approach to modelling implementations of charts.

\section{Implementations of Stateflow charts}
\label{sec:impl}

\begin{figure}[t]
\centering
\includegraphics[width=.8\textwidth]{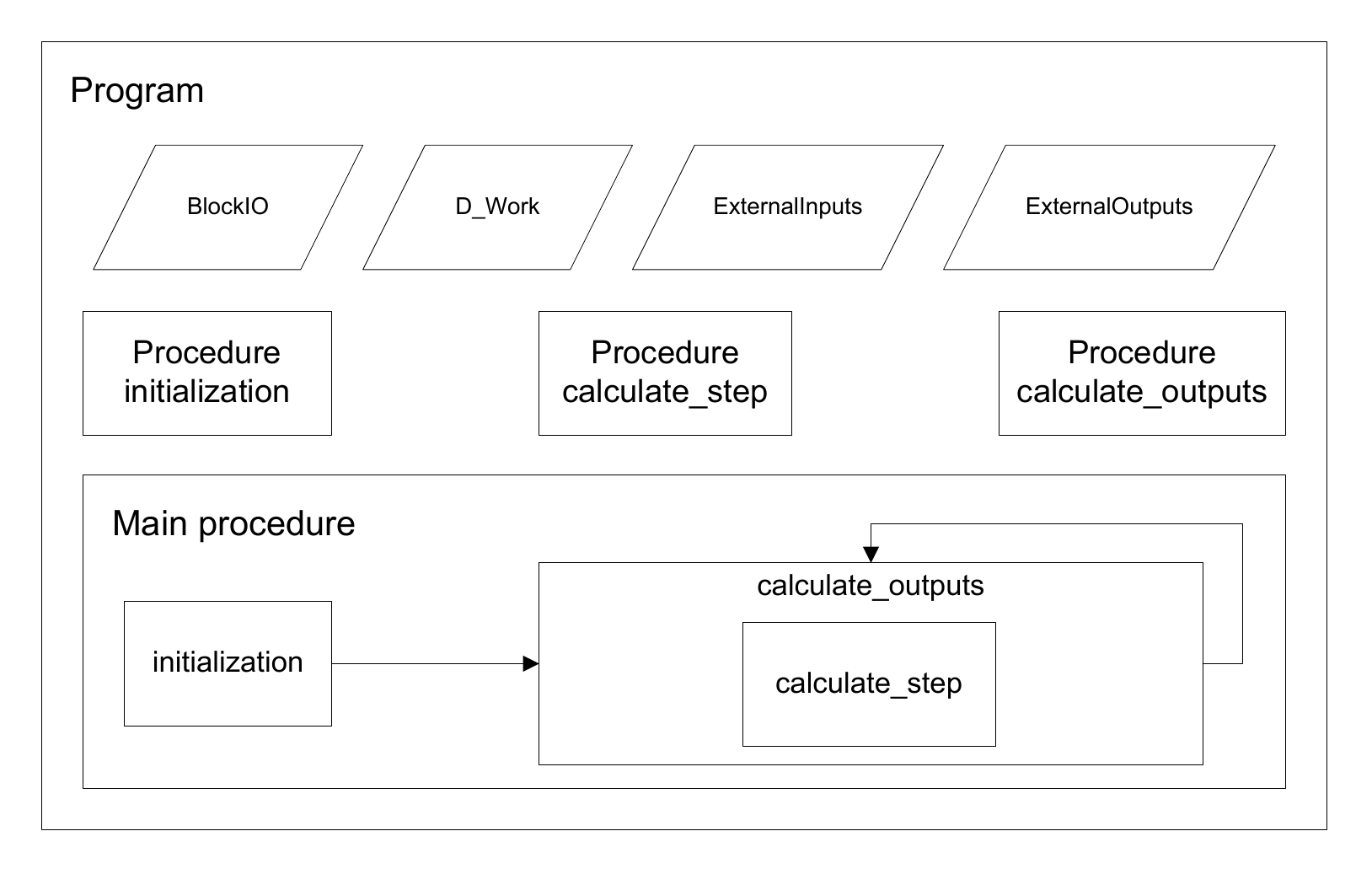}
\caption{Architcture of the implementations of Stateflow charts}
\label{fig:impl}
\end{figure}

Figure~\ref{fig:impl} shows the structure of implementations of Stateflow charts generated automatically by Real Time Workshop/Stateflow Coder. In general, the implementations produced consist of a series of structures that define the state of the chart (inputs, outputs, local variables, events, execution state, and so on) and a series of procedures. The procedures can be divided into those that implement the execution of the chart, which are relevant for our verification, and those that calculate the next time step. Since we capture time using synchronisation, we restrict our attention those of the first kind depicted in Figure~\ref{fig:impl} as \texttt{calculate\_outputs}, \texttt{initialization} and \texttt{calculate\_step}.

The procedure \texttt{calculate\_outputs} implements the execution of the chart, \texttt{initialization} initialises the variables of the program, and \texttt{calculate\_step} implements the control of the execution of the chart according to the number of active events. The main procedure of the implementation initializes the program and repeatedly calculate the outputs using the procedure \texttt{calculate\_step}. In the case of a C implementation, the procedures shown in Figure~\ref{fig:impl} are implemented as C functions whose names reflect the name of the chart. In our example, for instance, the procedure $calculate\_outputs$ is implemented as the C function \texttt{AbsoluteValue\_output}.

Of particular interest to us is how the implementation models information regarding the status of the states. This is done in two different ways, according to the type of
decomposition of the states. In this discussion, we regard the chart as a state. If the state has a parallel decomposition, its status is modelled by a single variable in the structure $D\_Work\_X$, where $X$ is the name of the chart. This variable is called $is\_active\_S$, where $S$ is the name of the state; it has a numerical type (\texttt{uint8\_T}), but, in fact, it is used as a boolean variable, that is, if its value is zero the state is not active, otherwise, it is active. 

If the state has a sequential decomposition, its status is modelled by two variables in $D\_Work\_X$. The variable $is\_active\_S$ is as described above. The variable $is\_S$ records a number that identifies which substate is active at the time, its value is zero if there are no active substates. If a state is a child of a state with a sequential decomposition, no variable of the form $is\_active\_S$ is created, as the information about its status is already recorded in the variable $is\_P$, where $P$ is the name of its parent state.%, otherwise the usual variables are defined for it.

In our example, we have only states with sequential decompositions. The status of the chart is recorded by \texttt{is\_active\_c1\_AbsoluteValue} and \texttt{is\_c1\_AbsoluteValue}. There is no need for variables that record the status of \texttt{P} and \texttt{N}, as this information can be obtained from \texttt{is\_c1\_AbsoluteValue}.

% In the cases where it is possible that two or more chart steps can be execute in one time-step, the chart step is implemented in a function \texttt{chartstep()}, and this function is called by \texttt{chart\_output()}. This situation occurs, for instance, when two or more edge-triggered events can trigger the execution of the chart at the same time (\cite{stateflow} p. 9-15).

%The function \texttt{chart\_initialize()} initialises some timing and logging information which we ignore, and initialises the chart data and control structures. The function \texttt{MDLInitialise} seems to redundantly and selective initialise the chart date and execution state. The documentation of Stateflow coder\cite{} and Real-Time Workshop\cite{} are not clear on the contents of these function in the context of Stateflow chart blocks. \emph{Look more}.

%\emph{Main characteristics of the implementations, in particular, the state of the program (inputs, outputs, and constrol structures), and the structure of the implementation ({\bf initialize} and {\bf output} functions).}

We model the implementation as a series of schemas and a single process. Figure~\ref{fig:impl-model} gives a partial view of the model of the implementation of our example. Schemas model the records in the implementation. For instance, \texttt{D\_Work\_AbsoluteValue} is modelled by the schema $D\_Work\_AbsoluteValue$.

\begin{figure}[t]
\centering
\hspace{-1.2cm}
\begin{minipage}{.9\textwidth}
\begin{circus}
D\_Work\_AbsoluteValue == [is\_active\_c1\_AbsoluteValue, is\_c1\_AbsoluteValue: \nat]\\
\ldots\\
\circprocess AbsoluteValue \circdef \circbegin\\
\circstate AbsoluteValue\_state == [AbsoluteValue\_DWork: D\_Work\_AbsoluteValue; \ldots]\\
AbsoluteValue\_DWork\_is\_c1\_AbsoluteValue \circdef x:\nat \circspot AbsoluteValue\_DWork :=\\
\t1 \lblot is\_active\_c1\_AbsoluteValue == AbsoluteValue\_DWork.is\_active\_c1\_AbsoluteValue,\\
\t1 is\_c1\_AbsoluteValue == \_y \rblot\\
\ldots\\
AbsoluteValue\_output \circdef tid: \num \circspot\\
\circblockbegin
\circblockbegin%3
\circif AbsoluteValue\_DWork.is\_active\_c1\_AbsoluteValue = 0 \circthen\\
AbsoluteValue\_DWork\_is\_active\_c1\_AbsoluteValue(1) \circseq\\
\circblockbegin%4
	\circif AbsoluteValue\_B.SineWave1 \geq 0 \circthen\\
		AbsoluteValue\_DWork\_is\_c1\_AbsoluteValue(AbsoluteValue\_IN\_P)\\
	\circelse \lnot (AbsoluteValue\_B.SineWave1 \geq 0) \circthen\\
		AbsoluteValue\_DWork\_is\_c1\_AbsoluteValue(AbsoluteValue\_IN\_N)\\
	\circfi
	\circblockend\\%4
\circelse \lnot (AbsoluteValue\_DWork.is\_active\_c1\_AbsoluteValue = 0) \circthen\\
\ldots\\
\circfi
\circblockend%3
\circseq\\
AbsoluteValue\_Y\_y(AbsoluteValue\_B.y)
\circblockend\\
\ldots\\
\circspot AbsoluteValue\_initialize \circseq \circmu X \circspot Input \circseq AbsoluteValue\_output\circseq Output \circseq end\_cycle \then X
\end{circus}
\end{minipage}
\caption{\Circus~model of the implementation of the chart in Figure~\ref{fig:example}}
\label{fig:impl-model}
\end{figure}

For each of the relevant C function in the implementation, we define a \Circus~action that models it. In Figure~\ref{fig:impl-model}, we present the \Circus~action that models the function
\texttt{AbsoluteValue\_output}. The main action of the process is fixed and consists of calling the initialisation action ($AbsoluteValue\_initialize$ for our example), and recursively reading the inputs (with the action $Input$), producing the outputs (with the action $AbsoluteValue\_output$), offering the outputs (with the action $Output$), and signalling the end of the ``time-step" by synchronising on $end\_cycle$. The actions $Input$ and $Output$ abstract as communications the shared variables used to implement inputs and outputs.

%Notice that the second to last line of code consists of an assignment from the chart output variable $y$ contained in the block \texttt{AbsoluteValue\_B} to the same variable in the block \texttt{AbsoluteValue\_Y}. This occurs because the new value of the output variable should only be available when the step has finished; this is reflected in our model. It is worth mentioning, that, in our model, if a block attempts to read an output of the chart before it is available, the block has to wait; this corresponds to the interpretation that the execution of blocks is determined by the wiring, and that one block will only be executed once all of its inputs are available.

In the modelling of the functions, we map C constructs to
similar \Circus~constructs. Loops are modelled using recursion. In general,
the translation of implementation constructs is direct, except for the assignment
to a variable of a structure. Since we cannot write $b.f := v$ (as a translation
of \texttt{b.f = v}), for a variable $b$ of type binding, we define \Circus~actions
that specify the assignment of a binding to the variable, as the action
$AbsoluteValue\_DWork\_is\_c1\_AbsoluteValue$ in Figure~\ref{fig:impl-model}. This action takes one parameter
$x$ of type $\nat$, and assigns a binding of the schema $D\_Work\_AbsoluteValue$
to the state component $AbsoluteValue\_DWork$. The binding is formed by associating each component
of the schema to a value, the component $is\_active\_c1\_AbsoluteValue$ is associated to the value of
the same component, and $is\_c1\_AbsoluteValue$ is associated to the value of the parameter.

In the next section, we discuss a refinement strategy that supports the verification of the models of implementations just described with respect to the models discussed in Section~\ref{sec:formal-model}.

%\emph{Discuss in general terms how to obtain the models of the implementation, and focus on the abstractions necessary: time, events, inputs and outputs.}

\section{Refinement Strategy}
\label{sec:strategy}

\begin{figure}
\centering
\includegraphics[width=\textwidth]{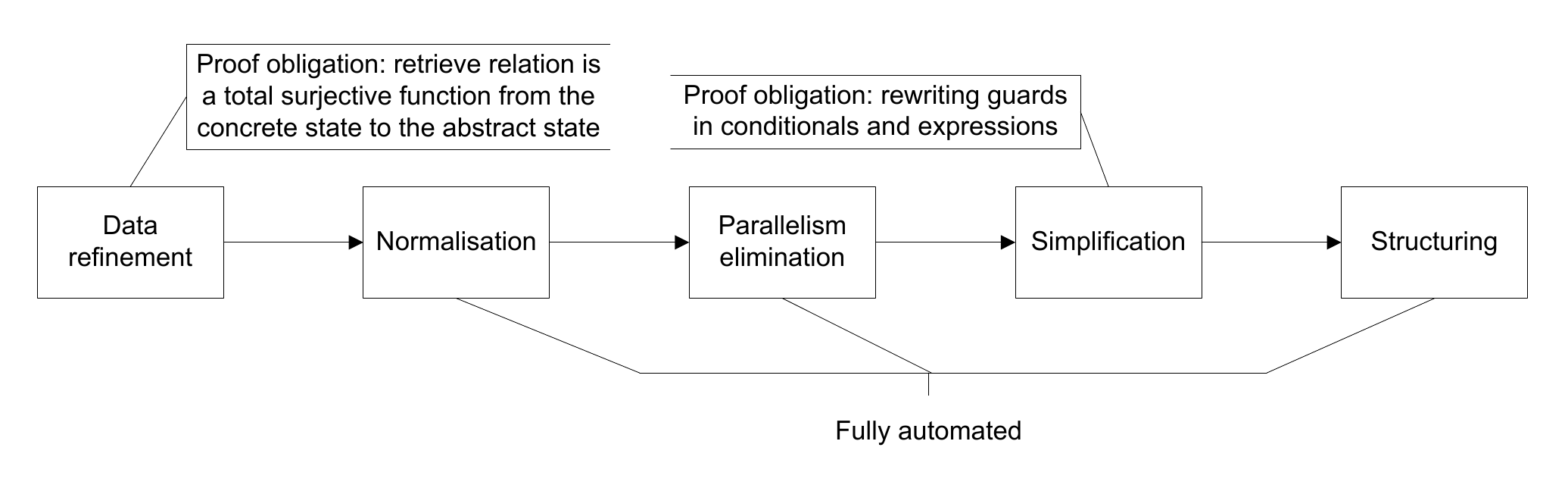}
\caption{Overview of the refinement strategy}
\label{fig:strategy}
\end{figure}

Our refinement strategy consists of five phases: data refinement, normalisation, parallelism elimination, simplification, and structuring. Figure~\ref{fig:strategy} illustrates the strategy; it identifies the fully automated phases and the proof obligations that stem from the other phases.

In the data refinement phase, we modify the state of the chart process in order to conform to the state of the implementation model. The normalisation phase transforms the parallel composition of the chart and simulation processes into a single process whose main action initialises the state and recursively offers an action that encodes a step of execution of the chart. The parallelism elimination phase collapses the parallel actions that occur in the resulting process. This is necessary because the parallelism in the diagram model reflects the operational semantics of Stateflow, not a parallel design for a program. In the simplification phase, we simplify expressions and predicates, and move assumptions through the model to eliminate unreachable branches of alternations. Finally, in the structuring phase, we rewrite the main action to match the functions of the implementation, and therefore the actions of its model.

The strategy proposed in this section is generic enough to be applied to a large class of charts and implementations. It takes advantage of restrictions on the architecture of the implementation to support a high degree of automation. We consider here only sequential implementations of Stateflow diagrams. The steps of the strategy are, however, useful in the refinement to parallel implementations as well.

In the sequel, we describe the details of each phase. They define procedures to apply existing and novel \Circus\ refinement laws whose soundness guarantees the soundness of our verification strategy.

\subsection{Data refinement}
\label{subsec:data}

In this phase, we construct a retrieve relation between the abstract state of the chart process and the concrete state of the implementation model. With that, we use the \Circus~calculus to construct a refinement of the chart process, and so preserve its structure, and transform the assignments, operation schemas, and communications. Precisely, we follow the procedure below for constructing retrieve relations that are total surjective functions from concrete to abstract states. They allow us to proceed with the data refinement in a calculational fashion.

\begin{figure}
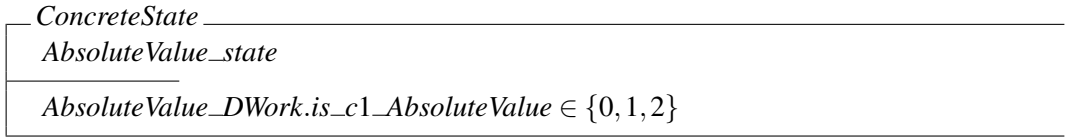

\begin{schema}{ConcreteState}
AbsoluteValue\_state
\where
AbsoluteValue\_DWork.is\_c1\_AbsoluteValue \in \{0,1,2\}
\end{schema}
\caption{Restricted concrete state}
\label{fig:concrete}
\end{figure}

The state components of the implementation model belong to one of four groups: execution specific data, like \texttt{AbsoluteValue\_DWork} in our example, local variables (like \texttt{AbsoluteValue\_B}), input variables (like \texttt{AbsoluteValue\_U}), and output variables (\texttt{AbsoluteValue\_Y}). Additionally, we can restrict the components of the concrete state to take values only over the appropriate sets. For example, Figure~\ref{fig:concrete} shows the state of our implementation model with one additional invariant; it requires that $is\_c1\_AbsoluteValue$ takes values from $\{0,1,2\}$.

The retrieve relation maps the execution specific data to the components of $SimulationData$, and the local, input and output variables to components of $SimulationInstance$. The correspondence between the input and output variables is trivial. It is obtained by equating the concrete variables to the abstract variable whose name is the same except for a prefix $v\_$. The specification of the relation between the execution specific data of the concrete state and the function $state\_status$ in $SimulationData$ is obtained by using a set comprehension where, for each state identifier $s$, we define a boolean $active$ that determines its status from the concrete state. These conditions can be calculated as follows. For a chart name $C$, and each component of $D\_Work\_C$ (in our example, the schema
$D\_Work\_AbsoluteValue$) named $is\_active\_name$ ($is\_active\_c1\_AbsoluteValue$, for example), where $name$ is the name of state (or chart), we have the following condition.
\[s = name \land active = (\IF~C\_DWork.is\_active\_name > 0 \THEN~\true~\ELSE~\false)\]
For instance, the condition for $is\_active\_c1\_AbsoluteValue$ equates $s$ to $c\_AbsoluteValue$, and $active$ to $\true$ or $\false$ depending on whether the value of $is\_active\_c1\_AbsoluteValue$ is greater than zero or not.

\begin{figure}
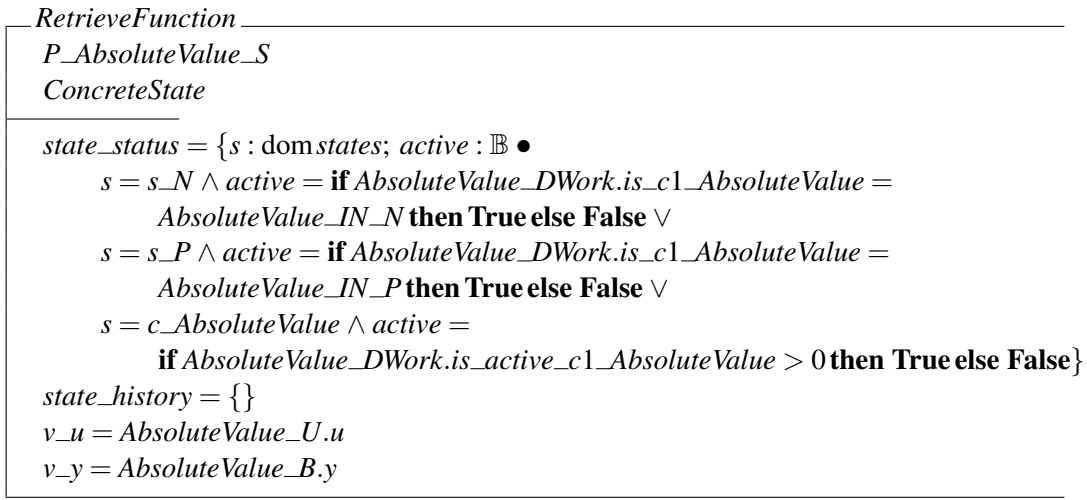

\begin{schema}{RetrieveFunction}
P\_AbsoluteValue\_S\\
ConcreteState
\where
state\_status = \{s: \dom states; active: \boolean \circspot\\\t1
s = s\_N \land active = \IF~AbsoluteValue\_DWork.is\_c1\_AbsoluteValue = \\
\t2 AbsoluteValue\_IN\_N \THEN \true \ELSE~\false \lor\\\t1
s = s\_P \land active = \IF~AbsoluteValue\_DWork.is\_c1\_AbsoluteValue =\\
\t2 AbsoluteValue\_IN\_P \THEN \true \ELSE~\false\lor\\\t1
s = c\_AbsoluteValue \land active =\\
\t2 \IF~AbsoluteValue\_DWork.is\_active\_c1\_AbsoluteValue > 0 \THEN~\true \ELSE~\false\}\\
state\_history = \{\}\\
v\_u = AbsoluteValue\_U.u\\
v\_y = AbsoluteValue\_B.y
\end{schema}
\caption{Total surjective functional retrieve relation}
\label{fig:retrieve}
\end{figure}

For each component of the schema $D\_Work\_C$ of the form $is\_name$, where $name$ is the name of a state (or chart), we formulate a condition in the following way. For each substate $X$ of $name$, we define the condition below.
\[s = s\_X \land active = (\IF~C\_DWork.is\_name = C\_IN\_X \THEN~\true~\ELSE~\false)\]
In our example, the condition for the state \texttt{P} equates $s$ to $s\_P$, and $active$ to $\true$ or $\false$ depending on whether the value of $is\_c1\_AbsoluteValue$ is $AbsoluteValue\_IN\_P$ or not. All these conditions are composed in a disjunction as shown in the definition of the retrieve relation for our example in Figure~\ref{fig:retrieve}.

%In the concrete state in Figure~\ref{fig:concrete}, a component $D\_Work\_AbsoluteValue$ of type $AbsoluteValue\_DWork$ is included by $AbsoluteValue\_state£$; it contains two components, $is\_active\_c1\_AbsoluteValue$ and $is\_c1\_AbsoluteValue$, where $is\_c1\_Absolutevalue$ is restricted to take values from the set \{0,1,2\}, where the value 1 corresponds to the name $AbsoluteValue\_IN\_P$, and the value 2 corresponds to the name $AbsoluteValue\_IN\_N$. 

Since our example does not contain history junctions, the implementation has no state components that model the state component $state\_history$ of the chart process. The model of the chart establishes that this state component is the empty partial function, therefore we equate $state\_history$ to the empty set.

The retrieve relation in Figure~\ref{fig:retrieve} is functional because each abstract state component is defined by a function of a component of the concrete state. Since no restriction is imposed on the concrete state for the applicability of the function, the relation is also total. Moreover, for every abstract state $A$, it is possible find a concrete state that is related to $A$ by the retrieve relation because the functions are invertible.

Using this retrieve relation, we apply the laws of simulation~\cite{Cavalcanti2003a,Oliveira2006a} to obtain a \Circus\ process $C\_P\_AbsoluteValue$ by data refinement of $P\_AbsoluteValue$, and to refine the process $AbsoluteValue$ to a process $CAbsoluteValue$, as shown in Figure~\ref{fig:data-refined-model}. We  define the constant $ss$ to increase the readability of $C\_P\_AbsoluteValue$. This function is defined as the characterisation of $state\_status$ in Figure~\ref{fig:retrieve}.

\begin{figure}
\centering
\hspace{-.8cm}
\begin{minipage}{\textwidth}
\begin{circus}
\circprocess C\_P\_AbsoluteValue \circdef \circbegin\\
\end{circus}
\vspace{-1.4cm}
\begin{axdef}
StateflowChart
\where
identifier = c\_AbsoluteValue\\
states = \{(s\_P, S\_P), (s\_N, S\_N), (c\_AbsoluteValue, C\_AbsoluteValue)\}\\
transitions = \{(t\_P\_N, T\_P\_N),  \ldots, (t\_6\_N, T\_6\_N)\}\\
junctions = \{(j5, J5), (j6, J6)\}
\end{axdef}
\vspace{-1cm}
\begin{axdef}
ss: SID \pfun \boolean
\where
ss = \{s: \dom states; active: \boolean \circspot\\\t1
s = s\_N \land active = \IF~AbsoluteValue\_DWork.is\_c1\_AbsoluteValue = AbsoluteValue\_IN\_N\\
\t2 \THEN~\true \ELSE~\false \lor\\\t1
s = s\_P \land active = \IF~AbsoluteValue\_DWork.is\_c1\_AbsoluteValue = AbsoluteValue\_IN\_P\\
\t2 \THEN~\true \ELSE~\false\lor\\\t1
s = c\_AbsoluteValue \land active = AbsoluteValue\_DWork.is\_active\_c1\_AbsoluteValue\}
\end{axdef}
\vspace{-1cm}
\begin{circusaction}
\circstate ConcreteState
\end{circusaction}
\vspace{-1.1cm}
\begin{zed}
CActivate == [\Delta ConcreteState; x?: SID | \ldots]\\
%CDeactivate == [\Delta ConcreteState; x?:SID | \ldots]\\
CInitState == [ConcreteState~' | AbsoluteValue\_DWork'.is\_active\_c1\_AbsoluteValue = \false \land \ldots]\\
\end{zed}
\vspace{-1.5cm}
\begin{circusaction}
%entryaction\_P \circdef executeentryaction.(s\_P)?o?ce \then \Skip\\
%\ldots\\
%duringaction\_P \circdef executeduringaction.(s\_P)?o?ce \then AbsoluteValue\_B\_y(AbsoluteValue\_U.u)\\
\ldots\\
condition\_P\_N \circdef \circif ((AbsoluteValue\_U.u \alt 0) \neq 0) \circthen \ldots
\ldots\\
%trigger\_P\_N \circdef checktrigger.(t\_P\_N)?e \then (result.(t\_P\_N).(e)!(\true) \then \Skip)\\
%\ldots\\
%getstate \circdef state?x\prefixcolon(x \in \dom(states))!(states(x)) \then \Skip\\
%\ldots\\
%activation \circdef activate?x \then \lschexpract CActivate \rschexpract\\
%\ldots\\
Inputs \circdef (read\_inputs \then (i\_u?x \then AbsoluteValue\_U\_u(x)))\\
Outputs \circdef (write\_outputs \then (o\_y!(AbsoluteValue\_B.y) \then \Skip))\\
%AllActions \circdef conditions \extchoice triggers \extchoice Inputs \extchoice Outputs \extchoice ChartActions \extchoice InterfaceActions\\
\circspot \lschexpract CInitState \rschexpract \circseq
\circmu X \circspot \circblockbegin\circblockbegin
\circmu Y \circspot
\circblockbegin
((AllActions\circinterrupt(interrupt\_chart\then\Skip))\circseq Y)\\
\extchoice\\
end\_cycle\then\Skip
\circblockend
\circblockend\circseq X\circblockend\\
\circend
\end{circusaction}
\vspace{-1cm}
\begin{circus}
\circprocess~CAbsoluteValue \circdef Simulator \lpar interface \cup \lchanset end\_cycle \rchanset \rpar C\_P\_AbsoluteValue
\end{circus}
\end{minipage}
\caption{Data refinement of the processes shown in Figure~\ref{fig:model}}
\label{fig:data-refined-model}
\end{figure}

The main action does not change, but components of the actions that it uses are transformed. For example, $Activate$ and $InitState$ are data refined to operate over the concrete state. Figure~\ref{fig:data-refined-model} shows part of the definition of $CInitState$ (the data refinement of the schema $InitState$); it shows the part of the predicate that defines the operation. The actions $condition\_P\_N$, $Inputs$, and $Outputs$ are also data refined; the first has the component $v\_u$ rewritten to $AbsoluteValue\_U.u$, the second has the assignment transformed into an action that assigns a value to a component of a schema binding (as mentioned in Section~\ref{sec:impl}), and the third has the component $v\_y$ substituted by $AbsoluteValue\_B.y$, in accordance with the retrieve relation.

In the next section, we describe how to collapse the parallel composition in $CAbsoluteValue$.

\subsection{Normalisation}

In this phase, we first collapse the parallelism between the chart and simulator processes in the process $CAbsoluteValue$, and rewrite the main action of the resulting new process to a normal form: an initialisation action, followed by a recursive action. This allows us to focus on the body of the recursion that characterises one step of execution.

We construct the new process by taking the state of the chart process (the simulator process is stateless), and combining the main actions of the chart and simulator processes in the same way the processes were combined, as shown in Figure~\ref{fig:main-action}. This is a direct application of the definition of the semantics of process parallelism in \Circus.

\begin{figure}
\centering
\begin{minipage}{\textwidth}
\begin{circusaction}
\circspot
\left(\begin{array}{c}
\circblockbegin
\lschexpract CInitState \rschexpract \circseq\\
\circmu X \circspot \circblockbegin
\circmu Y \circspot
\circblockbegin
\circblockbegin AllActions\\
\circinterrupt\\
(interrupt\_chart\then\Skip)\circblockend\circseq Y\\
\extchoice\\
end\_cycle\then\Skip
\circblockend\circblockend\circseq X
\circblockend\\
 \lpar \{AbsoluteValue\_B, AbsoluteValue\_DWork\} | interface \cup \lchanset end\_cycle \rchanset | \{\} \rpar\\
 (\circmu X \circspot Step \circseq X)\end{array}\right)\circhide interface
\end{circusaction}
\end{minipage}
\caption{Combined main action after merging the two processes.}
\label{fig:main-action}
\end{figure}

Next, we move the schema action $CInitState$ out of the parallel composition, distribute the hiding over the sequential composition, and eliminate the hiding over $CInitState$. The external recursion in the first action of the parallel composition, and the recursion in the second action are then merged. This is possible because $Step$ necessarily terminates in a synchronisation over the channel $end\_cycle$, since this channel is in the synchronisation set, and this synchronisation stops the inner recursion, and starts a new cycle of the external recursion. We are left with the action in Figure~\ref{fig:main-action-2}, which calls $CInitState$, and recursively executes the parallel composition of a recursion (the recursion over $Y$ in Figure~\ref{fig:main-action}), and the action $Step$, with the channels in $interface$ hidden.

%The phase of parallelism elimination starts by moving the schema action $InitState$ outside the parallel composition using the \emph{Schema/Parallel composition distribution} law presented in \cite{Oliveira2006a}, distribute the hiding using the law \emph{Hiding/Sequence distribution}~\cite{Oliveira2006a}, and eliminate the hiding from the schema action using the law \emph{Hiding identity}~\cite{Oliveira2006a}. Next, the external recursion in the first action of the parallel composition, and the recursion in the second action are merged using the law \emph{Recursion/Parallel composition distribution - 2}. This is possible because $Step$ necessarily terminates in a synchronisation over the channel $end\_cycle$, since this channel is in the synchronisation set, and this synchronisation stops the inner recursion, and starts a new cycle of the external recursion. We are with an action that calls $CInitState$, and recursively executes the parallel composition of a recursion (the recursion over $Y$ in Figure~\ref{fig:main-action}), and the action $Step$, with the channels in $interface$ hidden, as shown in Figure~\ref{fig:main-action-2}~\footnote{We abbreviate $interface$ to $if$ in Figure~\ref{fig:main-action-2}.}.

\begin{figure}
\centering
\begin{minipage}{\textwidth}
\begin{circusaction}
\circspot
\circblockbegin
\lschexpract CInitState \rschexpract \circseq\\
\circmu X \circspot 
\left(\begin{array}{c}
\circmu Y \circspot
\circblockbegin
\circblockbegin AllActions\\
\circinterrupt\\
(interrupt\_chart\then\Skip)\circblockend\circseq Y\\
\extchoice\\
end\_cycle\then\Skip
\circblockend\\
 \lpar \ldots \rpar\\
 Step\end{array}\right)\circhide interface\circseq X\circblockend
\end{circusaction}
\end{minipage}
\caption{Main action after the normalisation phase (We abbreviate the parallelism).}
\label{fig:main-action-2}
\end{figure}

\subsection{Parallelism elimination}

In this phase, we eliminate the parallelism still embedded in the main action. The parallel action inside the recursion of Figure~\ref{fig:main-action} reads an input event, reads the input variables, executes the chart, writes the outputs, and ends the cycle. In general, the step of execution involves a series of communications over channels that may or may not be in the synchronisation set. Communications over channels not in the synchronisation set are moved outside the parallel composition, and the others are used to select an action from $AllActions$ (or trigger an interruption). We proceed as follows to evaluate the communications and remove the parallelism.

By expanding the definition of $Step$ in Figure~\ref{fig:main-action-2}, we have two communications one after the other. The first is a communication over the channel $input\_event$ that is not in the synchronisation set, and the second is a synchronisation over $read\_inputs$, which is in the synchronisation set. We move the communication outside the parallel composition, unfold the recursion over $Y$, and resolve the communication on the channel $read\_inputs$. This produces a prefixing action identical to the body of the action $Inputs$ in Figure~\ref{fig:data-refined-model}. Because the communication is hidden, we eliminate it and obtain the action in Figure~\ref{fig:main-action-3}.

%For example, by expanding the definition of $Step$ in Figure~\ref{fig:main-action-2}, we have two communications one after the other. The first is a communication over the channel $input\_event$ that is not in the synchronisation set, and the second is a synchronisation over $read\_inputs$, which is in the synchronisation set. We first apply the law \emph{Input prefix/Parallelism composition - distribution 2}\cite{Oliveira2006a} and move the communication outside the parallel composition. Next, we apply the \emph{Recursion unfold} law~\cite{Cavalcanti2003a} to unfold the recursion, systematically apply the law \emph{Parallelism composition/External choice - distribution}\cite{Oliveira2006a} to distribute the parallelism the external choices in $AllActions$, and apply where possible the laws \emph{Parallelism composition Deadlocked 1}\cite{Oliveira2006a} and \emph{Communication/Parallelism composition - distribution}\cite{Oliveira2006a}. At this point we are left with an external choice where only one of the actions is different from $\Stop$, we then apply the law \emph{External choice - unit}\cite{Cavalcanti2003a} until we are left with a single action. Finally, the prefixing of the resulting action is simplified by applying the law \emph{Hiding Identity}.

\begin{figure}
\centering
\hspace{-1.4cm}
\begin{minipage}{\textwidth}
\begin{circusaction}
\circspot
\circblockbegin
\lschexpract CInitState \rschexpract \circseq\\
\circmu X \circspot \circblockbegin
input\_event?ie \then\\
\left(\begin{array}{c}
\circblockbegin
i\_u?x \then AbsoluteValue\_U\_u(x)\circseq\\
\circmu Y \circspot
\circblockbegin
\circblockbegin AllActions\circinterrupt\\(interrupt\_chart\then\Skip)\circblockend\circseq Y\\
\extchoice\\
end\_cycle\then\Skip
\circblockend\\
\circblockend\\
\lpar \ldots \rpar\\
\circblockbegin
\left(\begin{array}{c}
ExecuteChart(ie)\\
\circinterrupt\\
\circblockbegin
interrupt\_simulator \then\\ interrupt\_chart \then \Skip\circblockend\end{array}\right) \circseq\\
write\_outputs \then end\_cycle \then \Skip
\circblockend 
\end{array}\right)\circhide interface
\circblockend\circseq X
\circblockend
\end{circusaction}
\end{minipage}
\caption{Main action after the communications over $input\_event$ and $read\_inputs$ are treated.}
\label{fig:main-action-3}
\end{figure}

The actions that can be selected can be an atomic information request, as exemplified by $read\_inputs$, a non-atomic information request, or a Stateflow action request. In the first case, as already shown, the parallelism can be removed by resolving the communication. In the case of a non-atomic request (for instance, a trigger action of Figure~\ref{fig:model}), the action is composed of two communications. To eliminate the parallelism we resolve them; this potentially involves resolving a conditional expression that selects the appropriate value to communicate. A Stateflow action request consists of a communication that identifies the appropriate action, a series of \Circus~actions that encode the Stateflow action, and a synchronisation that indicates completion. In this case, the initial communication and the final synchronisation are treated as usual. The encoding of the Stateflow action contains assignments and local event broadcasts. Assignments are moved out of the parallel composition, and broadcasts produce a recursive execution of the chart, which can be treated using the same strategy.

The strategy for eliminating parallelism can be seen as a two level strategy. The first level is guided by the structure of the simulator process, which potentially leads to the execution of an action of the chart process, and the second level is guided by a chart process action that has been executed. Since the simulator process is the same for all charts, we explore its structure to define the refinement strategy.% as explained above.
The same is not true for the chart process, but due to the simple structure of the actions in the chart process, we can explore the limited patterns that occur.

At the end, we obtain a main action whose structure is as shown in Figure~\ref{fig:main-action-pre-final}. There, we have already simplified expressions, which is the objective of the next phase.

\subsection{Simplification}

In the simplification phase, we transform expressions and eliminate unreachable branches of conditional statements. The simplification of expressions takes advantage of the constants that model the structure of the chart, as well as state invariants. For instance, Figure~\ref{fig:main-action-part} contains an expression that appears frequently in communication resolutions: $states(c\_AbsoluteValue).identifier$. It evaluates to the identifier of the state whose identifier is $c\_AbsoluteValue$, but this is exactly $c\_AbsoluteValue$, thus we simplify it. 

\begin{figure}
\centering
\hspace{-1cm}
\begin{minipage}{\textwidth}
\begin{circusaction}
\left(\begin{array}{c}
\circblockbegin
\circmu Y \circspot
\circblockbegin
AllActions\circseq Y\\
\extchoice\\
end\_cycle\then\Skip
\circblockend\\
\circblockend\\
\lpar \ldots \rpar\\
\circblockbegin
status!(states(c\_AbsoluteValue).identifier)?active \then\\
\circblockbegin 
\circif active = \true \circthen\\\t1
\circblockbegin ExecuteActiveChart(states(c\_AbsoluteValue),ie)\circseq\\
write\_outputs \then end\_cycle \then \Skip\circblockend\\
\circelse active = \false \circthen\\\t1
\circblockbegin ExecuteInactiveChart(states(c\_AbsoluteValue),ie)\circseq\\
write\_outputs \then end\_cycle \then \Skip\circblockend\\
\circfi
\circblockend
\circblockend
\end{array}\right)\circhide interface
\end{circusaction}
\end{minipage}
\caption{Parallel action of the main action in Figure~\ref{fig:main-action-3} after further refinement steps.}
\label{fig:main-action-part}
\end{figure}

\begin{figure}[!h]
\centering
\hspace{-2.1cm}
\begin{minipage}{\textwidth}
\begin{small}
\begin{circus}
\circspot CInitState; \mu X \circspot input\_event?ie \then i\_u?x \then AbsoluteValue\_U\_u(x);\\
\circblockbegin 
\circif ss(c\_AbsoluteValue) = \false \circthen  \ldots\\
\circelse ss(c\_AbsoluteValue) = \true \circthen\\
\circblockbegin
\circif 
AbsoluteValue\_DWork.is\_c1\_AbsoluteValue = AbsoluteValue\_IN\_N \circthen\\
\circblockbegin
\circif ((AbsoluteValue\_U.u \ageq 0) \neq 0) \circthen\\
\circblockbegin
\circif 
AbsoluteValue\_DWork.is\_c1\_AbsoluteValue = AbsoluteValue\_IN\_P \circthen\\
Deactivate\_P;
Deactivate\_N;
Activate\_P;\\
\circblockbegin
\circif (AbsoluteValue\_U.u \alt 0) \neq 0) \circthen \ldots\\
\circelse \lnot (((AbsoluteValue\_U.u \alt 0) \neq 0)) \circthen \ldots\\
\circfi
\circblockend\\
\circelse 
AbsoluteValue\_DWork.is\_c1\_AbsoluteValue \neq AbsoluteValue\_IN\_P \circthen\\
Deactivate\_N;
Activate\_P;\\
\circblockbegin
\circif ((AbsoluteValue\_U.u \alt 0) \neq 0) \circthen \ldots\\
\circelse \lnot (((AbsoluteValue\_U.u \alt 0) \neq 0)) \circthen \ldots\\
\circfi
\circblockend\\
\circfi
\circblockend\\
\circelse \lnot (((AbsoluteValue\_U.u \ageq 0) \neq 0)) \circthen AbsoluteValue\_B\_y(\negate~AbsoluteValue\_U.u);\\
\circblockbegin
\circif ((AbsoluteValue\_U.u \alt 0) \neq 0) \circthen\\
\circblockbegin
\circif 
AbsoluteValue\_DWork.is\_c1\_AbsoluteValue = AbsoluteValue\_IN\_N \then\\\t1 \ldots\\
\circelse 
AbsoluteValue\_DWork.is\_c1\_AbsoluteValue \neq AbsoluteValue\_IN\_N \then\\\t1 \ldots\\
\circfi
\circblockend\\
\circelse \lnot (((AbsoluteValue\_U.u \alt 0) \neq 0)) \circthen \ldots\\
\circfi
\circblockend\\
\circfi
\circblockend\\
\circelse 
AbsoluteValue\_DWork.is\_c1\_AbsoluteValue \neq AbsoluteValue\_IN\_N \circthen\\
\circblockbegin
\circif ((AbsoluteValue\_U.u \alt 0) \neq 0) \circthen\\
\circblockbegin
\circif 
AbsoluteValue\_DWork.is\_c1\_AbsoluteValue = AbsoluteValue\_IN\_N \circthen \ldots\\
\circelse 
AbsoluteValue\_DWork.is\_c1\_AbsoluteValue \neq AbsoluteValue\_IN\_N \circthen \ldots\\
\circfi
\circblockend\\
\circelse \lnot (((AbsoluteValue\_U.u \alt 0) \neq 0)) \circthen \ldots\\
\circfi
\circblockend\\
\circfi
\circblockend\\
\circfi
\circblockend; X
\end{circus}
\end{small}
\end{minipage}
\caption{Partially simplified main action of process $AbsoluteValue$.}
\label{fig:main-action-pre-final}
\end{figure}

After these simplifications are carried, we obtain a main action as in Figure~\ref{fig:main-action-pre-final}. The last branch of the second conditional has the guard: $AbsoluteValue\_DWork.is\_c1\_AbsoluteValue \neq AbsoluteValue\small{\_IN\_N}$. Since $\small{AbsoluteValue\_IN\_N}\small{ = }2$, the invariant of the concrete state implies that
\[\hspace{-.6cm}AbsoluteValue\_DWork.is\_c1\_AbsoluteValue = 1 \lor AbsoluteValue\_DWork.is\_c1\_AbsoluteValue = 0\]
Therefore, since $AbsoluteValue\_IN\_P$ is a constant defined as 1, we replace the above guard with 
\[\hspace{-.6cm}AbsoluteValue\_DWork.is\_c1\_AbsoluteValue = AbsoluteValue\_IN\_P \lor\\
\hspace{-.6cm}AbsoluteValue\_DWork.is\_c1\_AbsoluteValue = 0\]
This guard is then broken in two, and a new branch is added to the conditional statement. We proceed in this way for every guard defined by a disjunction that checks the status of a state. This simplification is applied whenever the status of a state in a sequential decomposition is checked because our model always includes a branch whose guard is an inequality. It is necessary because, while our model contains only binary conditionals, the implementations may have conditionals with more than two branches.

We traverse the resulting action and for each conditional statement found, we attempt to simplify the guard. If we can simplify a guard to false, we eliminate the branch. If the guard is true, we reduce the whole conditional to the action it guards; this is correct, because all our conditionals are of the form $\circif b \circthen \ldots \circelse \lnot b \circthen \ldots \circfi$.
If a conditional statement cannot be simplified, it should match a conditional statement in the model of the implementation. For example, the first conditional statement in Figure~\ref{fig:main-action-pre-final} corresponds to the first conditional statement of the action $AbsoluteValue\_output$ in Figure~\ref{fig:impl-model}.\pagebreak

The fourth conditional statement in Figure~\ref{fig:main-action-pre-final} is an example of a statement that can be simplified. The guard of the first branch is $AbsoluteValue\_DWork.is\_c1\_AbsoluteValue = AbsoluteValue\_IN\_P$, but this is inside a branch whose guard is $AbsoluteValue\_DWork.is\_c1\_AbsoluteValue = AbsoluteValue\_IN\_N$, and since $is\_c1\_AbsoluteValue$ cannot have both values (and no statement modifies this component between the two branches), the first guard is false, and that branch can be eliminated. In this way, we put the model of the chart in the same shape as the model of the implementation, and once this is achieved, the structuring phase takes place. Formalisation of this strategy using refinement requires a law to introduce assumptions based on the guards of the conditional, laws to distribute and use assumptions, and finally a law to remove an assumption when it is no longer needed. These are standard laws that are valid in \Circus\ as shown in \cite{Oliveira2006a}.

\subsection{Structuring}

The structuring phase identifies each component of the main action that corresponds to an auxiliary action in the model of the implementation. It introduces this extra action in the process $CAbsoluteValue$, and uses the copy rule to replace the component of its main action by a call to it. The rationale behind this phase is to match the main action to that of the model of the implementation.

For instance, we compare the action $AbsoluteValue\_output$ in the model of the implementation to the subactions of the main action obtained from the previous phase.  We identify a subaction that is a match, introduce its definition, and substitute the name $AbsoluteValue\_output$ in the main action. The result of all this should be exactly the model of the implementation (as shown in Figure~\ref{fig:impl-model} for our example). If this is not the case, the verification has failed: either the program is wrong, or it does not conform to the architectural pattern that we can handle.

The detailed application of this strategy to our example can be found in \cite{Miyazawa2011b}.

\section{Conclusion}
\label{sec:conclusion}

We have proposed a refinement-based verification strategy for implementations of Stateflow charts. This strategy is guided by the structure of the models of Stateflow charts described in \cite{Miyazawa2011a}. %Moreover, we believe that the extension of the model of Stateflow charts to cover new features would not invalidate the strategy. 
We have also discussed how such a strategy can take advantage of the architecture imposed on generated code.

We have provided a procedure for obtaining retrieve relations that support the data refinement of the specification in a calculational style, thus rendering the data refinement phase also suitable for automation. In the case of the normalisation and parallelism elimination phases, the possibility of automation stems from the fixed structure of our models. The simplification phase
can be semi-automated because the main action consists of a number of nested if-statements, and assumptions generated by the guards of the conditional statements (among others) can be moved into the associated action, potentially falsifying some of the conditions in an internal conditional statement. Finally, the structuring phase can be guided by matching actions from the model of the implementation to subactions of the action being refined.

The refinement strategy for Simulink presented in
\cite{Cavalcanti2011} consists of four steps that systematically
collapse the massive parallelism of the diagram specification to
match the processes of the implementation model, prove that each of
the procedures in the implementation refine the action that specifies
it in the corresponding process, and finally, prove that the parallel
programs refine the process that specifies the system. The main
actions of component processes are put in a normal form where
they are defined as the iterative execution of a step that consists
 of reading the inputs in interleaving, calculating the
outputs and updating the state, writing the outputs in interleaving,
and synchronising on the channel $end\_cycle$.

Our strategy has a similar nature, however, it is worth mentioning some important differences. Our models
of Stateflow charts owe their parallelism to the separation between the structure of the model and the operational semantics of the simulator, not
to any implicit or explicit parallelism in the chart. Therefore, while in \cite{Cavalcanti2011} collapsing the parallelism
is guided by the implementation model, in our strategy it is performed until there are no parallel actions left. The
beginning of our parallelism elimination phase is similar to the process of putting the main action in a normal form
in the strategy for Simulink diagrams. The equivalent step in our model is, however, not as linear as in \cite{Cavalcanti2011}.
The decision to end the cycle comes from the action inherited from the simulator process. Nevertheless, by the end of
the parallelism elimination phase, we have a process whose main action is in the normal form of \cite{Cavalcanti2011}.
This suggests that not only our models can be integrated to the models of Simulink diagrams, but also that our strategy can be
used to put a Stateflow block in the normal form prescribed in \cite{Cavalcanti2011}. This will support the integrated
use of these verification techniques for Simulink diagrams involving Stateflow blocks.

There are several approaches to the formal analysis of Stateflow
diagrams. These works aim at the analysis of diagrams, not of their
implementations. Operational and denotational semantics are proposed
in~\cite{Hamon2004} and~\cite{Hamon2005};
these support static analysis, interpretation, and compilation of
Stateflow charts. Translations of Stateflow into notations that
support model checking are presented in~\cite{Banphawatthanarak2000},~\cite{Tiwari2002},~\cite{Scaife2004}, and~\cite{Chen2010}. Verification in these approaches is
based on temporal logics and bisimulation, rather than refinement,
thus verification of implementations is not the objective.
An approach based on Z to verify that
the chart satisfies a set of requirements of the system being
modelled is presented in \cite{Toyn2005}. However, it places strong restrictions on the Stateflow
notation.

Olderog \cite{Olderog1991} integrates three views of a system (trace specification,
process algebra and Petri nets) by formalising a relation between them. While this approach
ends in a graphical notation, namely Petri nets, we take the opposite direction: from a
graphical notation to a program. In~\cite{Bostrom2007}, the refinement calculus is adapted
to Simulink diagrams, but they do not cover Stateflow charts, and their goal is not the verification
of implementations, but the development of diagrams from contracts. In \cite{Reeve2006},
a semantics for $\mu$-Charts is constructed in Z, and a notion of refinement of $\mu$-Charts is derived from the existing Z refinement
calculus. This approach is similar to that presented in \cite{Miyazawa2011a}, where we define a semantics
of Stateflow charts in \Circus, thus allowing the \Circus\ refinement calculus to be applied, but it differs 
in the sense that we focus in a industrial non-formal notation, while the $\mu$-Charts notation is a simplification
of Statecharts mainly used in academia. Moreover, our approach goes beyond the
application of the refinement calculus to Stateflow charts; it addresses the problem of automation of the
refinement process.

As far as we know, this is the first work to address the issue of verification of implementations
of Stateflow charts. Moreover, as explained in detail~\cite{Miyazawa2011}, our models of Stateflow charts used as the base for the verification eliminate
many of the restrictions imposed in other formalisations.

Given the generality of our refinement strategy, we believe it scales
well to modified implementations. In particular, an implementation that does
not modify the use of the variable $AbsoluteValue\_DWork$ should still be amenable to
the specialisation of the refinement strategy discussed in Section~\ref{subsec:data}.
Moreover, our strategy can be used as a preliminary phase in the verification of parallel
implementations, as all the parallelism eliminated by the strategy
derives from the structure of the model.

As future work, we will address the issue of verification of parallel implementations of
Stateflow charts. Parallel implementations are not common, and as far as we know there are
no code generators that produce parallel implementations. We will extend the current strategy to allow
the verification to be carried out after the introduction of parallelism
in the implementation using fixed design patterns.

\bibliographystyle{eptcs}

\end{document}